\documentclass[twocolumn,10pt]{IEEEtran}
\usepackage{algorithm,algorithmic,amsbsy,amsmath,amssymb,epsfig,bbm,mathrsfs,multirow,amsthm}
\usepackage{tabu}
 \usepackage{booktabs}
\usepackage{multirow}
 \usepackage[table,xcdraw]{xcolor}
\usepackage{colortbl}
\usepackage{hyperref}
\usepackage{graphicx}
\usepackage{subfigure}

\DeclareMathAlphabet{\mathpzc}{OT1}{pzc}{m}{it}

\definecolor{mygray}{gray}{0.6}
\definecolor{myblue}{rgb}{0.8,0.85,1}

\begin{document}

\title{Securing Federated Learning: \\A Covert Communication-based Approach}
\author{Yuan-Ai~Xie, Jiawen~Kang, Dusit~Niyato,~\IEEEmembership{Fellow,~IEEE}, Nguyen~Thi~Thanh~Van, Nguyen~Cong~Luong*,  Zhixin~Liu, and Han~Yu\\
 \thanks{Yuan-Ai Xie and Zhixin Liu are with the Institute of Electrical Engineering, Yanshan University, Qinhuangdao 066004, China. Email:  xieyuan\_ai@163.com, lzxauto@ysu.edu.cn.}
 \thanks{Jiawen Kang is with School of Automation, Guangdong University of Technology. Email:kjwx886@163.com.}
 \thanks{Dusit Niyato, and Han Yu are with School of Computer Science and Engineering, Nanyang Technological University, Singapore 639798. Emails: dniyato@ntu.edu.sg, han.yu@ntu.edu.sg.}
 \thanks{Nguyen Thi Thanh Van is with the Faculty of Electrical and Electronic
Engineering, PHENIKAA University, Hanoi, Vietnam. Email: van.nguyenthithanh@phenikaa-uni.edu.vn.}
 \thanks{Nguyen Cong Luong* (Corresponding author) is with the Faculty of Computer Science, PHENIKAA
University, Hanoi, Vietnam. Email: luong.nguyencong@phenikaa-uni.edu.vn.}
}

\maketitle
\begin{abstract}
Federated Learning Networks (FLNs) have been envisaged as a promising paradigm to collaboratively train models among mobile devices without exposing their local privacy data. Due to the need for frequent model updates and communications, FLNs are vulnerable to various attacks (e.g., eavesdropping attacks, inference attacks, poisoning attacks, and backdoor attacks). Balancing privacy protection with efficient distributed model training is a key challenge for FLNs. Existing countermeasures incur high computation costs and are only designed for specific attacks on FLNs. In this paper, we bridge this gap by proposing the Covert Communication-based Federated Learning (CCFL) approach. Based on the emerging communication security technique of covert communication which hides the existence of wireless communication activities, CCFL can degrade attackers' capability of extracting useful information from the FLN training protocol, which is a fundamental step for most existing attacks, and thereby holistically enhances the privacy of FLNs. We experimentally evaluate CCFL extensively under real-world settings in which the FL latency is optimized under given security requirements. Numerical results demonstrate the significant effectiveness of the proposed approach in terms of both training efficiency and communication security.



\end{abstract}


\section{Introduction}
\label{sec:introduction}
Federated Learning Networks (FLNs) \cite{kairouz2021advances} have been proposed as a distributed privacy-preserving collaborative model training approach to alleviating societies' concerns on the exposure of sensitive data when building artificial intelligence applications. In FLNs, a server and a large number of mobile devices (MDs) perform multiple rounds of training iterations through wireless model updates to build machine learning models for specific tasks \cite{pmlr-v54-mcmahan17a}. 
However, FLNs with frequent model updates and communications are vulnerable to various types of privacy attacks, such as eavesdropping attacks, inference attacks, poisoning attacks and backdoor attacks, which in turn, have inspired a myriad of defense mechanisms (a.k.a. countermeasures) \cite{Lyu-et-al:2021}. 

However, there are still major limitations for existing countermeasures. On the one hand, each countermeasure is designed to address a specific attack. Hence, to tackle multiple different attacks, separate countermeasures are required. Such a defense strategy can be costly. On the other hand, existing countermeasures address attacks individually without unified security protection for FLNs. There are potential conflicts between some countermeasures when deployed together, which can further complicate FLN security issues. There is an urgent need for a unified, efficient, and highly secure solution to provide low-cost and effective defense for FLNs, and bring this field closer to real-world applications.

Recently, Covert Communication (CC) has been introduced as a promising security technique to prevent adversaries from detecting the existence of wireless transmission links \cite{lu2020intelligent,jiang2021covert,sobers2017covert}. Historically, the spread spectrum technique was adopted to achieve CC through spreading the transmitted signal power over a large time-frequency space. However, its covertness cannot be well analyzed. Hence, channel artifacts, such as additive white Gaussian noise channels, are used to hide communications. The fundamental information-theoretic limits of CC over random channels (i.e., the square root law) were explored in \cite{bash2013limits}. Specific CC techniques, such as Artificial Noises (AN) or jamming signals, were widely used to prevent attackers from detecting the legitimate transmissions \cite{lu2020intelligent,sobers2017covert}. 
Compared with the traditional cryptography and Physical Layer Security (PLS) technologies, CC can provide higher-level security by hiding transmissions that attract attackers' attention \cite{jiang2021covert}. 

As eavesdropping on the FL communication channels is often the first step for malicious third parties to launch attacks, we envision a \textit{Covert Communication-based Federated Learning (CCFL)} approach to secure the model transmissions from the MDs to the FL server. The following key technical challenge when re-contexting CC into FLNs needs to be resolved. When the jamming signals are introduced to help hide the model transmissions from the MDs to the server in FLNs, they inevitably degrade the transmission rates and thus increase the latency of FL. Hence, CCFL must jointly optimize the transmit power of each MD, the jamming power of the friendly jammer, and the local model accuracy at the MDs. This envisioned approach can contribute to the federated learning literature in the following ways:
\begin{enumerate}
\item \textbf{Holistic Security:} CCFL provides a unified and holistic security framework to mitigate a broad range of attacks on FL which involves eavesdropping on the communication channels between the MDs and the FL server. 
\item \textbf{Cost Effectiveness:} As CCFL is aimed at the key enabling step for malicious third parties to launch attacks on FLNs, it can preclude such attacks. In this way, MDs are no longer required to host computationally expensive countermeasures. This enables more resource-constrained devices to participate in FL.
\end{enumerate}


We carry out a case study grounded in real-world scenarios to showcase the potential benefits of CCFL, in which the AN-based CC technique is leveraged to hide the model transmissions from the MDs to the FL server from a warden (i.e., an attacker). Numerical results demonstrate significant advantages of the envisioned CCFL approach.

\section{Preliminaries}
\label{sec:FL}

\begin{figure*}[t!]
 \centering
\includegraphics[width=1\linewidth]{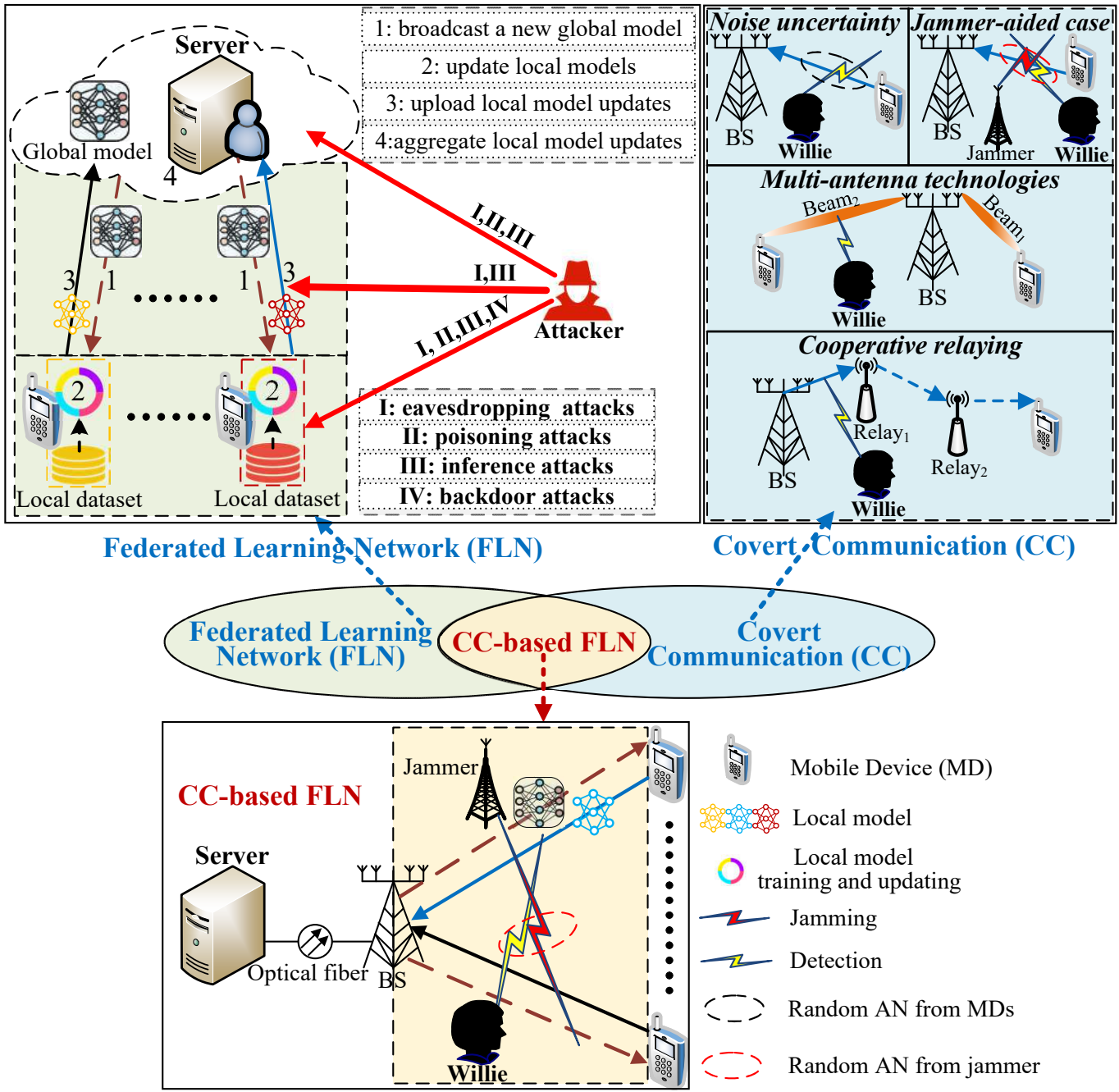}
 \caption{Covert communication-based model updates for secure FLNs.}
  \label{fig:1}
\end{figure*}


In a typical centralized learning network, the MDs are required to upload their local data to the central server through wireless links. Then, machine learning models are trained in the server (e.g., through Stochastic Gradient Descent (SGD) \cite{tan2020toward}). Nevertheless, the broadcast nature and limited spectrum of wireless networks as well as centralized data storage have led to critical issues including risks of privacy leakage, high communication overhead, and limited scalability.

To address the issues, FLNs have been proposed.
As shown in Figure \ref{fig:1}, the MDs obtain a shared global model broadcast by the FL server. They then train the local models with their data, and upload the local model parameters (e.g., gradients) to the server. After that, the server updates its global model by aggregating the received model updates (e.g., through federated averaging \cite{pmlr-v54-mcmahan17a}). These steps are repeated until the global model converges. During this process, the MDs transmit the model parameters instead of their local data. As a result, FLNs significantly reduces communication overheads and avoids privacy leakage by design.

\subsection{Attacks on FLNs and Countermeasures}
\label{sec:FL}
Due to the exposure of the communications between the MDs and the server to any interested and capable parties, FLNs generally have a large attack surface. Various attacks such as eavesdropping attacks, inference attacks, poisoning attacks, and backdoor attacks have been successfully mounted against FLNs \cite{Lyu-et-al:2021}. In the following part of this section, we present common attacks in FLNs and discuss the corresponding countermeasures.
\subsubsection{Eavesdropping Attacks}
In FLNs, the trained models can leak some sensitive information about the owners of the MDs (e.g., gender, occupation, and location) \cite{melis2019exploiting,lim2020federated}. In this context, eavesdropping attacks can occur when an adversary intercepts, deletes, or modifies these models that are transmitted between the FL server and the MDs. Since eavesdropping attacks are relatively easy to perform and can escalate to more severe cyber-attacks (e.g., Denial of Service (DoS)\footnote{DoS can overwhelm the server by making it go offline and deny further connection requests.} and jamming\footnote{Jamming can cause a poor model accuracy or even interrupt model transmissions.}), they are considered as one of the most common and fundamental attacks.
Based on whether the attacker listens to private conversations passively or actively, eavesdropping attacks can be categorized into passive and active eavesdropping attacks (a.k.a. man-in-the-middle attacks). Note that man-in-the-middle attackers pretend to be the intended FL server or MDs between these two entities in FLNs, and get access to control the traffic and fake the model transmissions. Unlike the passive eavesdropping attacks which are often regarded as less harmful, the man-in-the-middle attacks are severely harmful to FLNs.

\textbf{Countermeasures:} To mitigate eavesdropping attacks, cryptographic methods and PLS have been proposed. The cryptographic methods encrypt the transmitted models through a secret key that is only known by its intended receivers (e.g., the FL server). However, these methods also incur high computation costs and system complexity. This is especially challenging for FLNs involving a large number of MDs. 
Unlike the cryptographic methods, the main idea of PLS is to exploit the randomness of wireless channels and the AN (i.e., the jamming signal) to limit the quantity of models extracted or intercepted by an eavesdropper. Nevertheless, PLS cannot provide adequate security since the attackers are still able to capture part of the confidential FL model by side-channel analysis \cite{jiang2021covert}.

\subsubsection{Poisoning Attacks}
The two major types of poisoning attacks are data poisoning and model poisoning. Through modification of the training data (e.g., by flipping the labels randomly or specifically), the malicious MDs can launch data poisoning attacks and update incorrect model updates. Furthermore, the malicious MDs can flip the sign of benign model updates or adopt a predefined compromised model to craft poisoned model updates, which tamper with the FL model and reduces its performance \cite{tan2020toward}. Model poisoning can be regarded to include data poisoning, since data poisoning attacks ultimately act on the updated model, too.

\textbf{Countermeasures:} Two types of countermeasures are commonly employed to mitigate poisoning attacks on FLNs: 1) anomaly detection-based, and 2) robust aggregation. Anomaly detection-based methods are used to differentiate benign and poisoned model updates. For example, poisoned model updates can be identified and removed through analyzing their cosine similarities or mapped low-dimensional representations \cite{tan2020toward}. On the other hand, robust aggregation methods aim resist poisoned model updates. COMED, GEOMED, COTMED and KRUM are the commonly-used robust aggregation methods which replace the model averaging FL aggregation approach with component-wise median, geometric median, component-wise trimmed median and the shortest Euclidean distances from others, respectively \cite{tan2020toward}. In addition, these two types of countermeasures can be combined to into a workflow.
The main drawback is that they incur high computation costs, and are not suitable for deployment on MDs.



%

\subsubsection{Inference Attacks}
\label{sec:FL_sec}
Inference attacks fall largely into two categories: 1) membership inference attacks and 2) property inference attacks \cite{Lyu-et-al:2021}. The membership inference attacks aim to determine whether an exact data point was used to train a given model. By observing SGD-based gradient updates, attackers can infer a significant amount of private information and then may launch a powerful attack (e.g., gradient ascent attack) against other MDs.
On the other hand, property inference attacks aim to infer properties of training data that are independent of the characterized features of a class. Meanwhile, the attacker is assumed to have auxiliary training data correctly labeled with the property they intend to infer. 

\textbf{Countermeasures:} Differential Privacy (DP)-based solutions and encryption-based solutions are commonly used to mitigate inference attacks on FLNs. For DP-based solutions, a rigorous randomization mechanism (e.g., a Gaussian noise mechanism), is designed to inject additive noises into the trained parameters before they are uploaded to the FL server \cite{lim2020federated}. This guarantees that the addition or removal of a single data sample or model parameter does not affect the outcome of any inference. For example, \cite{abadi2016deep} introduced a differentially private SGD algorithm that can effectively protect the privacy of parameters trained by deep neural networks. However, due to the added noise in the local models from the MDs, the overall model accuracy suffers. Encryption-based solutions leverage encryption techniques to secure the data privacy of the MDs when the local model parameters are shared. On this basis, \cite{aono2017privacy} proposed a homomorphic encryption-based technique, which can protect sensitive information while preserving model performance. Nevertheless, they incur high computation costs and require complex system designs.

\begin{table*}
\centering
\caption{Summary of various high-risk attacks on federated learning networks (FLNs) and the corresponding countermeasures}\label{T1}
\resizebox*{1\linewidth}{!}{
\begin{tabular}{|c|c|c|c|c|}
\hline
\rowcolor[HTML]{CBCEFB}
Attack types          & Attack effect             & Source of Vulnerability                                                                 & Countermeasures                                                                                             & Limitations                                                                                \\ \hline
Eavesdropping attacks &  \begin{tabular}[c]{@{}c@{}}Medium \\ but prevailing\end{tabular} & \begin{tabular}[c]{@{}c@{}}MDs, compromised server\\wireless transmission \end{tabular} & Cryptographic methods/PLS                                                                                   & \begin{tabular}[c]{@{}c@{}}High computation costs/\\ inadequate security\end{tabular}              \\\hline
Poisoning attacks     & High                  & MDs, compromised server                                                                  & \begin{tabular}[c]{@{}c@{}}Anomaly detection-based  methods/\\ robust  aggregation  methods\end{tabular}    & High computation costs                                                                  \\ \hline
Inference attacks     & High                  & MDs, compromised server                                                                  & \begin{tabular}[c]{@{}c@{}}Differential privacy-based protection/\\ encryption-based solutions\end{tabular} & \begin{tabular}[c]{@{}c@{}}Relatively poor model accuracy/\\ high computation costs\end{tabular}             \\ \hline
Backdoor attacks      & High                  & MDs                                                                                     & Pruning and fine-tuning                                                                                     & Low-rank security                                                                      \\ \hline
\end{tabular}
}
\end{table*}

\subsubsection{Backdoor Attacks}
Backdoor attacks aim to inject
a malicious task into the FL model without affecting the
performance of the model on the actual learning task. Compared with poisoning and inference attacks, backdoor attacks are more subtle. Hence, the backdoor attacks are significantly more challenging to detect, especially when the accuracy of the model on the intended learning task does not show any variation which can alert the users to investigate the causes.

\textbf{Countermeasures:} There are two commonly used defenses against backdoor attacks: 1) pruning and 2) fine-tuning. By eliminating neurons that are dormant on clean inputs, pruning reduces the size of a compromised network, thereby disabling backdoor behaviors. However, a stronger pruning-aware attack can be mounted to evade pruning-based defense by concentrating the clean and backdoor behavior on the same set of neurons. Hence, \cite{liu2018fine} proposed the fine-tuning defense which retrains a small number of local models on a clean training dataset. Nevertheless, neither of these countermeasures offer adequate protection against backdoor attackers at the moment.

Existing countermeasures can protect FLNs against the corresponding attacks to different extents. However, as summarized in Table \ref{T1}, all countermeasures have some limitations when addressing these diverse attacks. They often incur high computation costs or decrease the efficiency of FLN training, which makes deployment on MDs challenging. 
Moreover, in the worst-case scenario in which multiple types of attacks are launched simultaneously, the detection of such attacks and deployment of countermeasures might result in unintended complications in addition to the prohibitively high resource requirements and system complexity. Thus, a unified, efficient, and highly secure FLN protection framework is needed for this technology to become widely adopted. 

\section{Covert Communication-based Federated Learning}
\label{sec:FL_sec}

\subsection{Covert Communication}
\label{sec:sec:FL_vul}
Covert Communication (CC), a.k.a. low probability of detection (LPD) communication, aims to mask the existence of a legitimate wireless transmission from a watchful adversary under the requirement of a certain covert rate for the intended user \cite{lu2020intelligent,bash2013limits}. Generally, CC provides three major advantages. Firstly, different from PLS which prevents an adversary from knowing the messages sent by the transmitter, CC prevents an adversary from knowing whether the transmission has occurred. If the adversary cannot detect the transmission, it will be unable to launch further attacks. Secondly, unlike encryption technologies, CC is low-cost, and its performance does not depend on the adversary's computation capability. Thirdly, CC has wide compatibility and can be easily adopted to complement advanced distributed artificial intelligence techniques, such as FL.


To improve the covertness of wireless links, various approaches have been developed based on the CC technique \cite{lu2020intelligent}.
\begin{enumerate}
    \item \textit{Noise Uncertainty:} Two major sources of uncertain noises are leveraged: background noise and random AN \cite{lu2020intelligent}. The background noise is easily affected by environmental factors such as temperature and humidity. Hence, an appropriate communication scenario or time should be chosen to enhance the covertness of the communication link. Conversely, the random AN can flexibly and efficiently amplify interference dynamics and confuse the adversary \cite{lu2020intelligent}. In contrast to background noise, the random AN is highly controllable and can be designed to exhibit different distributions, which greatly improves communication link covertness.
    \item \textit{Multi-Antenna Technologies:} By adequately exploiting spatial degree of freedom, multi-antenna technologies can help enhance the covertness of wireless links from all directions \cite{zheng2019multi}. Its realization requires beamforming to generate spatial selectivity. Specifically, a beamformer adjusts the corresponding amplitude and phase of the signals on each element of an antenna array in such a way that the superimposed radiation pattern is constructive in the desired direction and destructive in other directions. As a result, the transmitted signals can reach the desired receiver to enhance the data rate and simultaneously null the transmission at the adversary site. As the number of antennas increases, the antenna array will have a higher beamforming resolution which can be utilized to achieve a more reliable covert rate.
    \item \textit{Jammer-Aided Technologies:} There can be two types of friendly jammers for enhancing the stealthiness: 1) scheduled jammers, and 2) random jammers \cite{sobers2017covert}. A scheduled jammer is informed about the transmission of the legitimate transmitter and releases its AN with optimized parameters (e.g., jamming power). In contrast, a random jammer is unaware of the transmission by the legitimate transmitter and rather randomly or continuously transmits its AN. Compared with the random jammer, the scheduled jammer is more efficient and has more reliable covertness performance. For example, at the time the legitimate transmitter starts to transmit a codeword, the scheduled jammer turns down the power of the transmitted Gaussian noise, which is turned back up at the moment the transmitter finishes transmitting.
    \item \textit{Cooperative Relaying:} By leveraging the cooperation from intermediate node(s), cooperative relaying can achieve CC \cite{lu2020intelligent}. Note that the access distance shows a significant effect on covertness. For long-distance communication, high transmit power is required to achieve a target rate, which unavoidably impairs the covertness. To remedy this issue, multi-hop forwarding-based cooperative relaying is used. The fundamental is to shorten the communication distance of each hop to maintain the required transmit power low, leading to a low detection probability of the adversary. Through this technique, the covertness performance can be considerably enhanced.
\end{enumerate}

CC has demonstrated its superiority. It is a promising foundation for building a holistic security framework for FLNs. To achieve this goal, there are still several technical challenges that need to be resolved. On the one hand, the unified security of CC may not be enough to counter some attacks, especially when the adversary possesses strong detection capability. Hence, it is necessary to introduce additional covertness techniques. On the other hand, the enhancement of system covertness may be at the expense of other system performance metrics, such as latency or transmission rate. This motivates us to focus on the overall resource optimization of FLNs and forge a well-functioning system. In the next section, we discuss a vision towards a covert communication-based federated learning (CCFL) framework.

\label{sec:FL_sec}
\subsection{The Envisioned CCFL Approach}
In FLNs, the learning process involves multiple rounds of communications between MDs and the FL server. Adversaries can launch eavesdropping attacks and extract model parameter information via a weak channel condition. By detecting the existence of model updates, it is possible for eavesdropping attacks to escalate into more severe forms of attacks (e.g., DoS, jamming, and black holes) to manipulate FL model updates and aggregations.

To curb these cyber attacks effectively and preclude them from escalating, we envision incorporating CC into the FLN training process to hide the occurrence of model update transmissions. Without awareness of these transmissions, it is hard for the adversary to lunch attacks effectively. 
Since the large-scale FL devices are always configured with orthogonal channels, CC can deliver holistic security for the FL MDs through distributedly deploying it to each orthogonal channel. In addition, the distributed power control for CC incurs lower computation costs than existing countermeasures such as cryptographic methods and PLS. 
Multi-antenna technologies and cooperative relaying are more suitable for the downlink global FL model broadcast. The noise uncertainty and jammer-aided techniques are useful for securing the more vulnerable uplink FL model updates.

\subsection{A Case Study of CCFL}
We show a case study in which the jammer-aided technology is used to secure the local models transmitted from the MDs to the FL server for defending against eavesdropping attacks.

\subsubsection{System Model}
\label{sec:FL_sec}
\begin{figure}[h]
 \centering
\includegraphics[width=9cm]{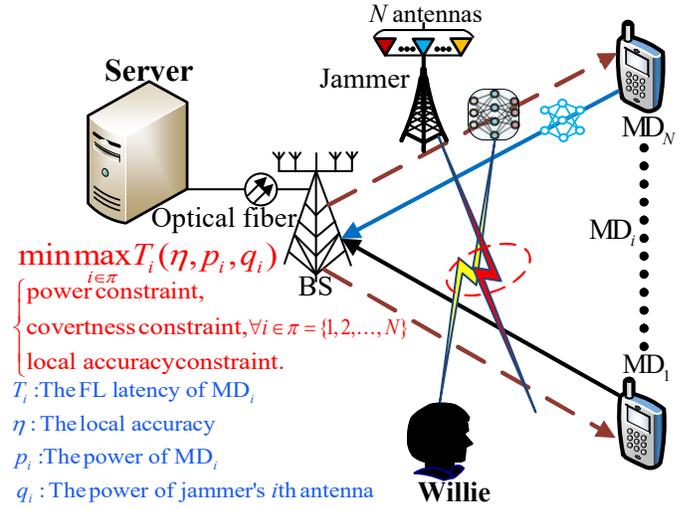}
 \caption{Jammer-based covert communication for FLNs.}
  \label{Covert_Communication_MCML}
\end{figure}

We consider an FL network as shown in Figure~\ref{Covert_Communication_MCML} which consists of $N$ devices, one FL server located at a Base Station (BS), and an attacker Willie. To achieve high efficiency, the orthogonal frequency-division multiple access technique is used for local model uploading by the MDs. A friendly jammer with $N$ antennas is deployed to transmit AN signals continuously with total power $p^j$ to Willie. To secure the model transmissions of all the devices, the jammer leverages the barrage jamming technique that transmits the AN signals over the full bandwidth occupied by the devices. Note that the AN signals also cause interference to the BS and may reduce the Signal-to-Interference-plus-Noise-Ratio (SINR) at the BS. The jammer is self-interested and rational. Thus, the server needs to pay the jammer a fee for the jamming service. For simplicity of discussion, a linear cost model is adopted in which the cost paid to the jammer is linearly proportional to $p^j$.

The FL model training process involves multiple iterations. In each iteration, the MDs train their local models to achieve a local accuracy $\eta$ (in terms of training error). At the end of each iteration, each MD can decide to transmit or not to transmit its local model to the FL server with a pre-defined transmission probability. When a device decides to transmit its model update to the server, and Willie judges that the device does not execute the transmission, then a \textit{miss-detection} occurs. When Willie judges that the device is transmitting the update while the device does not, then a \textit{false alarm} occurs. We define the covert probability for a device as the sum of the false alarm probability and the miss detection probability. We expect a high covert probability for situations in which Willie cannot correctly detect the model transmission of any device in the network. For this, the cover probability for the device in the FL network needs to be greater than a security requirement $(1-\epsilon)$~\cite{shahzad2019covert}, where $\epsilon$ is the security threshold. This is the CC constraint.

\begin{figure*}[t!]
\centering
  \subfigure[FL latency versus the number of devices $N$]{\includegraphics[width=0.47\linewidth]{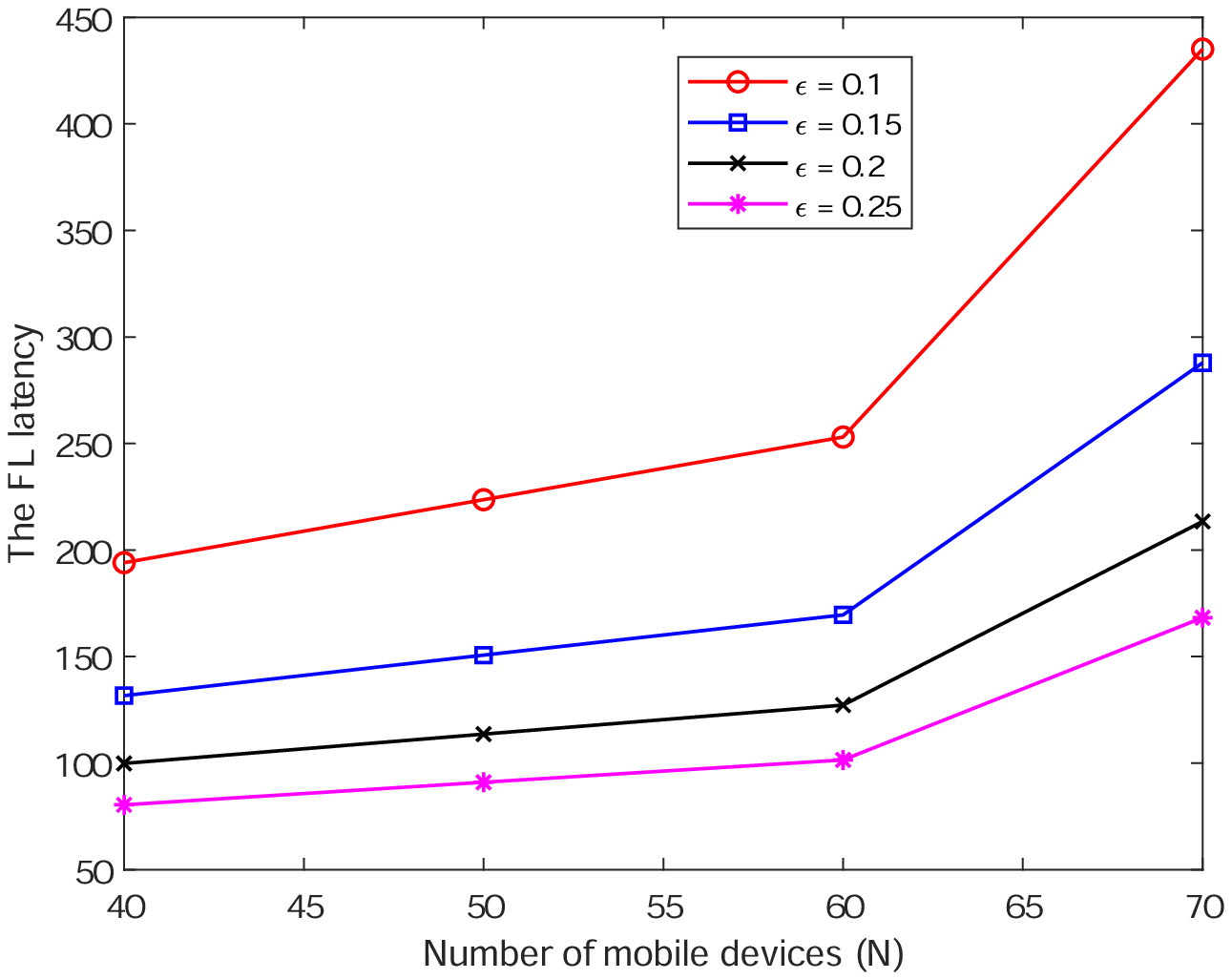}\label{CCFL_change:a}}
  \subfigure[FL latency and security performance versus security threshold $\epsilon$]{\includegraphics[width=0.51\linewidth]{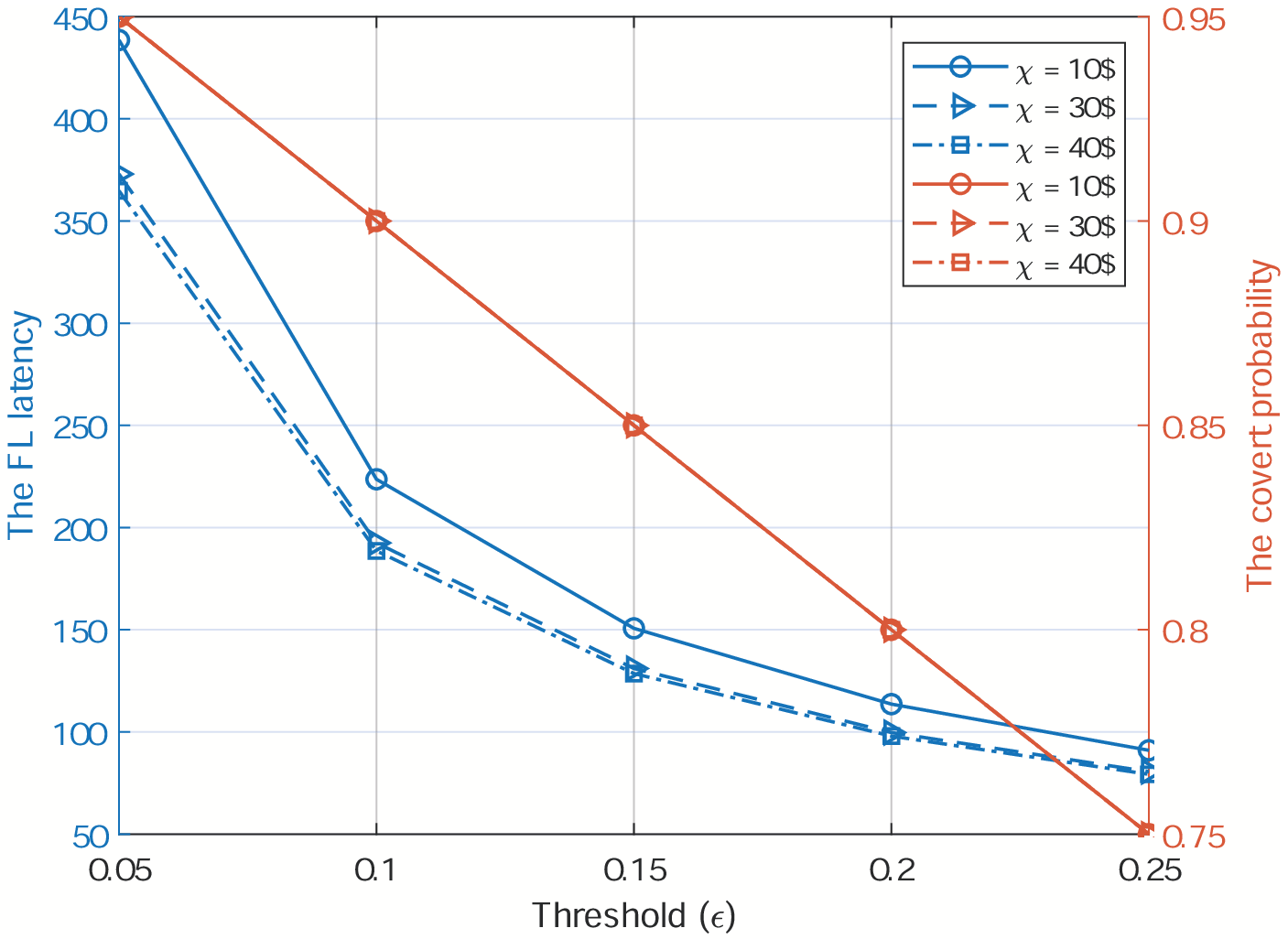}\label{CCFL_change:b}}
  \caption{The number of devices is set as $N=50$. The devices, jammer, and Willie are distributed randomly in a square area of size of $500$ m $\times$ $500$ m. The transmission probability of the devices is $0.7$, and their maximum power is $10$ dBm. Each device has $500$ data samples for its local training, and the device has a maximum computation capacity of $2$ GHz. The total bandwidth for the model transmissions of the devices is $20$ MHz. The price per jamming power unit is $0.5$\$ that is set by the jammer, and the budget of the server is $30$\$. The security requirement is $\epsilon=0.1$.}
  \label{CCFL_change}
\end{figure*}

To prevent Willie from detecting the local model transmissions by the devices, the server can request the jammer to transmit the AN signal with a higher power. However, this increases the cost that the server needs to pay the jammer and also reduces the SINR at the BS, leading to an increase in FL latency. Otherwise, the server can increase the local accuracy $\eta$ at the devices to reduce the number of local iterations, thereby reducing the computation time at the devices. However, this approach requires more global iterations to achieve high accuracy, thereby increasing the FL latency. Therefore, the joint optimization problem must simultaneously consider: 1) jamming power, 2) transmit power of the devices, and 3) local accuracy at the devices, in order to minimize the FL latency, subject to: 1) the CC constraint, 2) the maximum power of the devices, 3) the maximum power of the jammer, and 4) the FL server's budget. Here, the FL latency is defined as the maximum latency among the devices. The objective function and the CC constraint are non-convex. Thus, the optimization problem is non-convex. To solve the problem, we can adopt an alternating descent algorithm. The algorithm divides the original problem into two sub-problems that are alternately optimized at each iteration using successive convex approximation.

\subsubsection{Numerical Results}
\label{sec:FL_sec}




This part discusses the impact of important parameters on the latency and the security performance of the FL network.
Figure~\ref{CCFL_change:a} shows the impact of the number of MDs $N$ and the security threshold $\epsilon$ on FL latency. As can be observed, given the security threshold $\epsilon$, the FL latency increases as the number of devices $N$ increases. The reason is that the same fixed bandwidth is allocated to more devices. This decreases the transmission rate of each device, thereby increasing FL latency. It can also be observed from Figure~\ref{CCFL_change:a} that, as the security threshold $\epsilon$ increases, FL latency decreases. The reason is that as $\epsilon$ increases, the security requirement $(1-\epsilon)$ decreases. The low-security requirement allows the devices to transmit the models with higher transmit power. This leads to increases in SINR at the BS, thereby decreasing FL latency. Recall that when $(1-\epsilon)$ decreases, Willie can detect the transmissions of the devices more easily. As such, there is a trade-off between security performance and FL training efficiency.

Now, we discuss the impacts of the security threshold $\epsilon$ and the server's budget $\chi$, FL latency and security performance. As shown in Figure~\ref{CCFL_change:b}, as $\epsilon$ increases, the covert probability of the FL network decreases. This is obvious since as $\epsilon$ increases (i.e., $(1-\epsilon)$ decreases), a lower covert probability is enough to satisfy the CC constraint. Figure~\ref{CCFL_change:b} further shows that the covert probability remains almost constant over different budget settings by the FL server. The reason is that the covert probability depends on the ratio of the jamming power to the transmit power of the devices. As the budget varies, the jamming power bought from the jammer and the transmit power committed by the devices change together to satisfy the security requirement. Therefore, the covert probability remains unchanged over diverse budget values.

Nevertheless, varying the budget of the server leads to changes in FL latency. As shown in Figure~\ref{CCFL_change:b}, as we decrease the server's budget from $\chi=\$30$ to $\chi=\$10$, FL latency increases. The reason is that the server with a low budget can only buy a low amount of power from the jammer. The lower jamming power requires the devices to reduce their transmit power in order to satisfy the security requirement $(1-\epsilon)$ (i.e., to prevent Willie from detecting the transmissions from the MDs). This leads to decreases in SINR at the BS, thereby increasing FL latency. Under a higher budget setting (i.e., $\$30$), FL latency does not change significantly. The reason is that the server already finds an optimal jamming power that minimizes the FL latency while guaranteeing the security requirement, and it does not need to buy more power from the jammer. For this, the devices are not allowed to increase the transmit power due to the fixed security requirement. Thus, FL latency remains stable even when the budget is high.
\section{Conclusions and Future Directions}
\label{sec:}
In this article, we present a vision towards the holistic and cost-effective protection of FLNs from attacks through a covert communication-based approach. We start by discussing existing security issues for distributed collaborative training of FL networks. We then review key existing covert communication techniques, and present a case study in which jammer-based CC is used to prevent an attacker from detecting the local model transmissions by the devices involved in an FL setting. The use of the AN signals leads to the increase of the FL latency. Thus, we have investigated the FL latency minimization problem subject to the CC constraint. The numerical results provide an overview of the impact of important parameters such as the number of devices, security requirement, and budget of the server on FL latency and security performance. To the best of our knowledge, this is the first exploration on the potential of leveraging CC to enhance the security of FLNs.

For this emerging field of research, many interesting and challenging problems remain open:
\begin{itemize}
\item \textit{Impact of multiple attackers:} In this work, we assume that there is a single attacker in the FL network. In fact, there may be multiple attackers, and the server needs to prevent all of them from detecting the transmissions of the devices. This is challenging since the attackers may have different detection capabilities. One solution is to design the server to focus on defending against the attacker with the best detection capability.
\item \textit{Dynamic pricing of jamming power:} In this work, the jamming power price is fixed. In fact, the jammer can be self-interested and can dynamically set the price in different FL iterations to maximize its benefit. The server needs to account for time-varying prices to adapt the jamming power to avoid exceeding its budget.
\item \textit{Use of intelligent reflection surface (IRS):} To reduce the high cost for the jamming power, IRS can be deployed to enhance CC in the FL network. An IRS consists of reconfigurable reflecting elements that can reshape the phases, amplitudes, and
reflecting the angles of the environmental signals. For this, the phase shifts of the IRS can be configured to maximize the SINR at the BS, subject to the CC requirement.
\end{itemize}



\section*{Acknowledgments}
This research is supported by the National Research Foundation, Singapore under its AI Singapore Programme (AISG Award No: AISG2-RP-2020-019); the Nanyang Assistant Professorship (NAP); and the RIE 2020 Advanced Manufacturing and Engineering (AME) Programmatic Fund (No. A20G8b0102), Singapore. Any opinions, findings and conclusions or recommendations expressed in this material are those of the author(s) and do not reflect the views of National Research Foundation, Singapore.

\bibliographystyle{IEEEtran}
\bibliography{FLCovert}


\end{document}